\documentstyle[aps,epsf]{revtex}
\begin{document}
\draft
\title{Interaction-Free Measurement in mesoscopic systems and the reduction 
postulate}

\author{S.A. Gurvitz}

\address{Department of Particle Physics, Weizmann Institute of
         Science, Rehovot 76100, Israel}

\date{\today}

\maketitle

\begin{abstract}
We show that a noninvasive,``negative-result measurement'' can be realized in 
quantum dot systems. The measurement process is studied by applying 
the Schr\"odinger equation to the  whole system (including the detector). 
We demonstrate that the possibility of 
observing a particular state out of coherent superposition leads 
to collapse of the corresponding nondiagonal density-matrix elements
of the measured system. No additional reduction postulate is needed. 
Experimental consequences of the collapse time and the 
relativistic requirement are discussed for mesoscopic and optical systems.    
\end{abstract}
\twocolumn
According to the principles of quantum mechanics, the density-matrix 
of a system in a linear superposition 
of several states collapses
to the statistical mixture after measurement,    
$\sum_{nm}|n\rangle\rho_{m,n}\langle 
m|\to\sum_{m}|m\rangle\rho_{m,m}\langle m|\,$. This is the
von Neumann projection postulate in terms of the density-matrix 
formalism\cite{neu}. Since both the measuring device (the {\em detector}) 
and the measured system are described by the Schr\"odinger equation, 
the question arises of how such a non-unitary process takes place. 
It was suggested that the dissipative interaction of 
a measured system with the detector might 
be responsible for the density-matrix 
collapse\cite{zur}. Yet, such an explanation would not be valid if 
the detector does not distort the measured system, for instance, for 
``negative result'' measurements, when the information on the 
measured system is obtained by {\em non-observing} it at a given 
place\cite{renninger}. In this case the density-matrix collapse should be 
generated by pure ``informative'' process. It leads to 
another question of how fast such a collapse takes place and whether 
it can be accommodated with the relativistic requirement.   
  
A weak point of many studies of the measurement problem is the lack 
of a detailed quantum mechanical treatment of the entire 
system, that is, of the detector and the measured system together. 
The reason is that 
the detector is usually a macroscopic system, the quantum mechanical 
analysis of which is very complicated. Mesoscopic systems 
might be more useful for study of the measurement problem. 
We thus consider measurement of an electron in coupled quantum dots\cite{vdr} 
by a detector showing single electron charging  
of a quantum dot\cite{pep}. We demonstrate how the ``negative result''
measurement can be designed in this case.  
Since the entire system is rather simple for detailed quantum mechanical 
treatment,  the collapse of the density-matrix due to such 
an interaction-free measurement and its influence on the resonant current 
can be followed in great detail.  Although the investigated system  
looks rather specific, it bears all essential physics of the measurement 
process. The method can be also used for analysis of different systems,  
in particular for optical ``negative result'' measurements\cite{kwiat}.
 
We start with description of a measurement of  
single electron charging of a quantum dot, which eventually determines the 
resonant current, flowing through this dot. The system       
is shown schematically in Fig. 1\cite{pep}. The detector (the upper dot) is 
in close proximity to the lower dot (the measured system). 
Both dots are coupled to two separate reservoirs 
at zero temperature. The resonant levels $E_0$ and $E_1$ are below 
the corresponding Fermi levels, $\tilde E_L^F$, $E_L^F$ in the left 
reservoirs. 
In the absence of electrostatic interaction between electrons, 
the dc resonant currents in the detector and  
the measured system are respectively\cite{bg} 
\begin{equation}
I_D^{(0)}=e\frac{\gamma_L\gamma_R}{\gamma_L+\gamma_R},
~~~~~~~I_S^{(0)}=e\frac{\Gamma_L\Gamma_R}{\Gamma_L+\Gamma_R},
\label{a1}
\end{equation}
where $\gamma_{L,R}$ and $\Gamma_{L,R}$ are the tunneling 
partial widths of the levels 
$E_0$ and $E_1$ respectively, due to coupling with left and right 
reservoirs. The situation is different in 
the presence of electron-electron interaction between the dots, 
$H_{int}=Un_0n_1$, 
where $n_{0,1}$ are the occupancies of the upper and the lower dots
and $U$ is the Coulomb repulsion energy.  
If $E_{0}+U>\tilde E_F^L$, an electron from the left 
reservoir cannot enter the upper dot when the lower dot is 
occupied [Fig.~1~(b)]. 
On the other hand, if an electron occupies the upper dot
[Fig.~1~(a$^{\prime}$,b$^{\prime}$)], the displacement of the level 
$E_1$  of the lower dot is much less important, since it remains below 
the Fermi level, $E_{1}+U< E_F^L$. The upper dot can thus be considered as a detector,  
registering electrons entering the lower dot\cite{pep}. 
For instance, by measuring the detector current, 
$I_D$, one can determine the current in the lower dot, $I_S$. 
\begin{figure}
\vspace{0.2cm}
\hspace{0.4cm}
\epsfxsize=8.2cm
\epsfysize=8.2cm
\epsffile{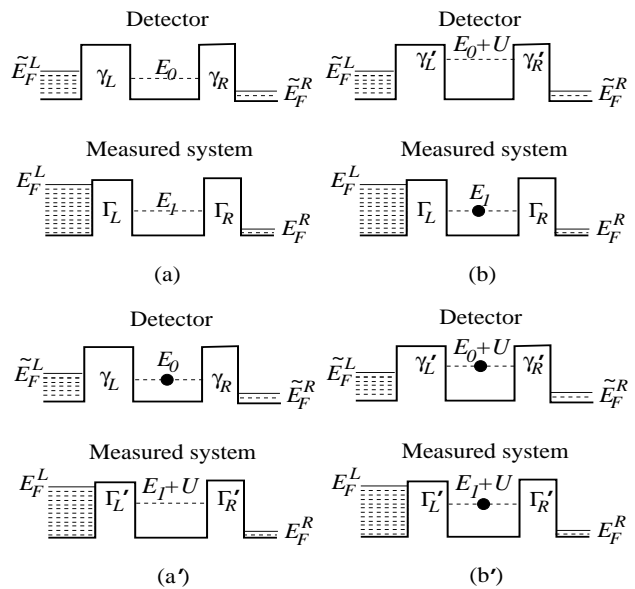}
\caption{The measurement of resonant current in a single-dot structure by 
another, nearby dot. 
All possible electron states of the detector (the upper well)
and the measured system (the lower well) are shown. Also indicated are 
the tunneling rates ($\gamma$ and $\Gamma$) of the detector and the 
measured system respectively.}
\label{fig1}
\end{figure}

Actually, the described detector affects the measured system. 
It happens when the detector is occupied 
[Fig.~1~(a$^{\prime}$,b$^{\prime}$)]. In this case an electron 
enters the lower dot at the energy $E_1+U$. As a result, the corresponding 
tunneling rates are modified ($\Gamma_{L,R}\to \Gamma'_{L,R}$), 
so that the measured current is distorted. One finds, however, 
that the states with empty detector, 
$|a\rangle$ and $|b\rangle$ [Fig.~1~(a,b)], 
do not distort the measured system. Nevertheless, the 
measurement process does take place:   
the detector current is 
interrupted whenever an electron occupies the measured system, but it 
flows freely when the measured system is empty. Such a measurement
is in fact the {\em negative} result
measurement\cite{renninger,kwiat}.  
Therefore, in order to transform the above detector to a 
noninvasive one, we need to diminish 
the role of the states 
$|a'\rangle$ and $|b'\rangle$ in the measurement process.
It can be done by varying penetrability of the detector barriers,   
$\gamma_R\gg \gamma_L$, so that  
the dwelling time of an electron inside the detector decreases.    
(Indeed, the average charge inside 
double-barrier structure\cite{bg} 
$e\gamma_L/(\gamma_L+\gamma_R )\to 0$ 
for $\gamma_L/\gamma_R\to 0$). In this case   
an electron entering the detector leaves it immediately, 
remaining the measured system undistorted.   

Now we confirm it explicitly by direct evaluation of the measured 
current that the above measurement is a noninvasive one.  
The currents through the detector and the measured system are 
determined by the density-matrix for the entire system $\rho (t)$,
which obeys the Schr\"odinger equation 
$i\dot\rho (t)=[{\cal H},\rho ]$ for ${\cal H}=H_D+H_S+H_{int}$, 
where $H_{D,S}$ are the tunneling Hamiltonians of the
detector and the measured system, respectively, and 
$H_{int}=U_in_0n_1$. 
The current in the detector (or in the measured system) 
is the time derivative of the total average charge $Q(t)$ accumulated
in the corresponding right reservoir (collector): $I(t)=\dot Q(t)$, 
where $Q(t)=e$Tr$[\rho^R(t)]$ and $\rho^R(t)$ is the 
density-matrix of the collector. 
It was shown\cite{glp,gp} that $I(t)$ is directly related to the 
density-matrix of the multi-dot system $\sigma (t)$,
obtained from the total density-matrix $\rho (t)$ by tracing out 
the reservoir states. 
One finds that the current in the detector or in the 
measured system is given by 
\begin{equation}  
I(t)=e\sum_c\sigma_{cc}(t)\Gamma^{(c)}_R, 
\label{a4}
\end{equation}
where $\sigma_{cc}\equiv \langle c|\sigma |c\rangle$
and the sum is over states $|c\rangle$ in which the well 
adjacent to the corresponding collector is occupied. 
$\Gamma_R^{(c)}$ is the partial width of the state $|c\rangle$
due to tunneling to the collector  
($\gamma_R$ or $\Gamma_R$). 
In turn, $\sigma (t)$ obeys the following system of the rate 
equations \cite{gp}
\begin{mathletters}
\label{a2}
\begin{eqnarray}
&&\dot\sigma_{aa}  = -(\gamma_L+\Gamma_L)\sigma_{aa}
+\gamma_R\sigma_{a'a'}+\Gamma_R\sigma_{bb}
\label{a2a}\\
&&\dot\sigma_{bb}  = -\Gamma_R\sigma_{bb}+\Gamma_L\sigma_{aa}
+(\gamma'_L+\gamma'_R)\sigma_{b'b'}
\label{a2b}\\
&&\dot\sigma_{a'a'}  = -(\gamma_R+\Gamma'_L)\sigma_{a'a'}+\gamma_L\sigma_{aa}
+\Gamma'_R\sigma_{b'b'}
\label{a2c}\\
&&\dot\sigma_{b'b'}  = -(\gamma'_L+\gamma'_R+\Gamma'_R)\sigma_{b'b'}
+\Gamma'_L\sigma_{a'a'},
\label{a2d}
\end{eqnarray}
\end{mathletters}
where the states $|a\rangle$, $|b\rangle$, $|a'\rangle$, $|b'\rangle$
are the available states of the entire system, Fig. 1. 

The currents in the detector and in the lower dot are    
$I_D(t)/e=\gamma_R\sigma_{a'a'}(t)+\gamma'_R\sigma_{b'b'}(t)$
and $I_S(t)/e=\Gamma_R\sigma_{bb}(t)+\Gamma'_R\sigma_{b'b'}(t)$,
respectively, Eq.~(\ref{a4}). 
The stationary (dc) current corresponds to $I=I(t\to\infty )$. 
Solving Eqs. (\ref{a2}) in the limit $\gamma_R,\gamma'_R\gg
\gamma_L,\gamma'_L$ we find 
\begin{equation}
\frac{I_D}{I_S}=\frac{\gamma_L}{\Gamma_L}\;,
~~~~~~~~~~~
I_S=e\frac{\Gamma_L\Gamma_R}{\Gamma_L+\Gamma_R}=I_S^{(0)}\;.  
\label{a5}
\end{equation}
The first equation shows that one can measure $I_S$ by measuring 
the detector current $I_D$,   
no matter how low the current $I_S$ is   
(providing that the ratio $\gamma_L/\Gamma_L$ is large enough).
On the other hand, the current $I_S$ is not distorted 
by the detector, as follows from the second equation. 

Consider now resonant transport in the coupled dot
structure shown in Fig. 2, where the upper dot is the detector. 
For simplicity, we assume strong Coulomb repulsion 
inside the coupled-dot, so only one electron can occupy  
it\cite{naz}. $U_{1,2}$ is the Coulomb repulsion
energy between the detector and the measured system when an 
electron occupies the first or the second dot of the measured system.
Similar to the previous case, Fig.~1,  each of the states,
$|a\rangle$, $|b\rangle$, $|c\rangle$ has its 
counterpart $|a'\rangle$, $|b'\rangle$, $|c'\rangle$ (not shown in Fig.~2)
corresponding to the occupied detector. 
We consider $E_0+U_1> \tilde E_F^L$, but $E_0+U_2< \tilde E_F^L$.
It means that 
the detector is blocked only when an electron occupies the first dot.
We assume that   
$U_2$ is very small so that the corresponding tunneling rates 
for entering the dots are not modified.
\begin{figure}
\vspace{0.2cm}
\hspace{0.4cm}
\epsfxsize=6.2cm
\epsfysize=8cm
\epsffile{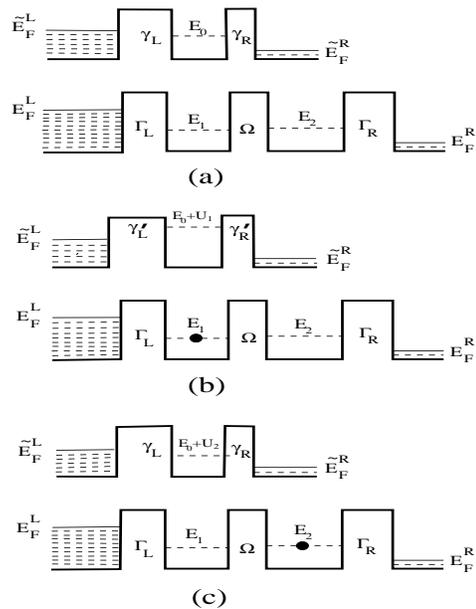}
\caption{The measurement of resonant current in a double-dot structure.
Only the states with empty detector (the upper dot) are shown.}
\label{fig2}
\end{figure}

Consider first the case of no detector (the upper dot is 
``switched off"). The available states of the double-dot system, 
$|a\rangle$, $|b\rangle$ and $|c\rangle$, are those  
as shown in Fig.~2. 
The (non-distorted) resonant current through this system 
is described by the modified rate equations, which in fact are 
the ``optical'' Bloch equations\cite{gp,naz}
\begin{mathletters}
\label{a3}
\begin{eqnarray}
&&\dot\sigma_{aa}  = -\Gamma_L\sigma_{aa}+\Gamma_R\sigma_{cc}
\label{a3a}\\
&&\dot\sigma_{bb}  = \Gamma_L\sigma_{aa}+i\Omega (\sigma_{bc}-\sigma_{cb})
\label{a3b}\\
&&\dot\sigma_{cc}  = -\Gamma_R\sigma_{cc}-i\Omega (\sigma_{bc}-\sigma_{cb})
\label{a3c}\\
&&\dot\sigma_{bc}  = i\epsilon\sigma_{bc}+i\Omega (\sigma_{bb}-\sigma_{cc})
-\frac{1}{2}\Gamma_R\sigma_{bc},
\label{a3d}
\end{eqnarray}
\end{mathletters}
where $\epsilon =E_2-E_1$ and $\sigma_{cb}=\sigma_{bc}^*$.  
The diagonal density-matrix elements $\sigma_{ii}$ are 
the probabilities of finding the system 
in one of the states, $|a\rangle$, $|b\rangle$ 
and $|c\rangle$. In the distinction with  
resonant tunneling through a single dot, 
the diagonal density-matrix elements are coupled with the 
non-diagonal elements $\sigma_{bc}$, $\sigma_{cb}$ (``coherences'')
that provide the hopping between two isolated 
levels, $E_1$ and $E_2$\cite{gp}. 
The total resonant dc flowing through 
this system is $I_S^{(0)}=e\Gamma_R\sigma_{cc}(t\to\infty )$, Eq.~(\ref{a4}).
One obtains\cite{naz}  
\begin{equation}
I^{(0)}_S=e\frac{\Gamma_R\Omega^2}{\epsilon^2+\Gamma^2_R/4
+\Omega^2(2+\Gamma_R/\Gamma_L)}
\label{a7}
\end{equation}
Note that the dissipation of the ``coherences'', $\sigma_{bc}$,  
is generated by the last term in
Eq.~(\ref{a3d}), proportional to the half of decay rates of 
the states $|b\rangle$ and  $|c\rangle$ due to their coupling with 
the reservoirs. (In our case the state $|b\rangle$ cannot decay, 
but only the state $|c\rangle$, so    
the dissipation is proportional to $\Gamma_R$). Since 
the resonant current proceeds via hopping between two dots, 
generated by $\sigma_{bc}$, it should decrease with dissipation:   
$I_S^{(0)}\to 0$ for $\Gamma_R\to\infty$, Eq.~(\ref{a7}).  

Now we ``switch on''the detector. 
The available states of the entire system are 
$|a\rangle$, $|b\rangle$, $|c\rangle$, Fig. 2, and 
$|a'\rangle$, $|b'\rangle$, $|c'\rangle$, corresponding to 
empty and occupied detector, respectively. The rate equations
describing the transport in the entire system are\cite{gp}  
\begin{mathletters}
\label{a6}
\begin{eqnarray}
&&\dot\sigma_{aa} =-(\Gamma_L+\gamma_L)\sigma_{aa}+\gamma_R\sigma_{a'a'}
+\Gamma_R\sigma_{cc}
\label{a6a}\\
&&\dot\sigma_{a'a'} =-(\Gamma'_L+\gamma_R)\sigma_{a'a'}
+\gamma_L\sigma_{aa}+\Gamma_R\sigma_{c'c'}
\label{a6b}\\
&&\dot\sigma_{bb} =\Gamma_L\sigma_{aa}+i\Omega (\sigma_{bc}-\sigma_{cb})
+(\gamma'_L+\gamma'_R)\sigma_{b'b'}
\label{a6c}\\
&&\dot\sigma_{b'b'} = \Gamma'_L\sigma_{a'a'}+i\Omega' (\sigma_{b'c'}-\sigma_{c'b'})-(\gamma'_L+\gamma'_R)\sigma_{b'b'}
\label{a6d}\\
&&\dot\sigma_{cc} = -(\Gamma_R+\gamma_L)\sigma_{cc}-i\Omega  (\sigma_{bc}-\sigma_{cb})+\gamma_R\sigma_{c'c'}
\label{a6e}\\
&&\dot\sigma_{c'c'} = -(\Gamma_R+\gamma_R)\sigma_{c'c'}-i\Omega'  (\sigma_{b'c'}-\sigma_{c'b'})+\gamma_L\sigma_{cc}
\label{a6f}\\
&&\dot\sigma_{bc} = i\epsilon\sigma_{bc}+i\Omega (\sigma_{bb}-\sigma_{cc})
\nonumber\\
&&~~~~~~~~~~~~~~~~~-\frac{1}{2}(\Gamma_R+\gamma_L)\sigma_{bc}
+\frac{1}{2}(\gamma_R+\gamma'_R)\sigma_{b'c'}\\
\label{a6g}
&&\dot\sigma_{b'c'} = i(\epsilon-U_1+U_2)\sigma_{b'c'}
+i\Omega' (\sigma_{b'b'}-\sigma_{c'c'})\nonumber\\
&&~~~~~~~~~~~~~~~~~~~~~~~~~~
-\frac{1}{2}(\Gamma_R+\gamma_R+\gamma'_L+\gamma'_R)\sigma_{b'c'}
\label{a6h}
\end{eqnarray}
\end{mathletters}
Let us again take the limit of the negative result measurement,  
$\gamma_R, \gamma'_R\gg\gamma_L,\gamma'_L$, when the detector is  
expected not to influence the measured system. If so,  
the density-matrix of the entire system, 
traced over the detector states would coincide 
with the density-matrix for the double-dot system without detector, 
Eqs.~(\ref{a3}).
We thus introduce   
$\bar\sigma_{aa}=\sigma_{aa}+\sigma_{a'a'}$,
$\bar\sigma_{bb}=\sigma_{bb}+\sigma_{b'b'}$, $\bar\sigma_{cc}=\sigma_{cc}+\sigma_{c'c'}$, and 
$\bar\sigma_{bc}=\sigma_{bc}+\sigma_{b'c'}$, 
which is the density matrix traced over the detector states.    
One finds from Eqs.~(\ref{a6}) that $\bar\sigma_{ij}$ obeys the 
following equations
\begin{mathletters}
\label{a8}
\begin{eqnarray}
&&\dot{\bar\sigma}_{aa}  = -\Gamma_L\bar\sigma_{aa}+\Gamma_R\bar\sigma_{cc}
\label{a8a}\\
&&\dot{\bar\sigma}_{bb}  = \Gamma_L\bar\sigma_{aa}+i\Omega (\bar\sigma_{bc}-\bar\sigma_{cb})
\label{a8b}\\
&&\dot{\bar\sigma}_{cc}  = -\Gamma_R\bar\sigma_{cc}-i\Omega (\bar\sigma_{bc}-\bar\sigma_{cb})
\label{a8c}\\
&&\dot{\bar\sigma}_{bc}  = i\epsilon\bar\sigma_{bc}+i\Omega (\bar\sigma_{bb}-\bar\sigma_{cc})
-\frac{1}{2}(\Gamma_R+\gamma_L)\bar\sigma_{bc},
\label{a8d}
\end{eqnarray}
\end{mathletters}

Let us compare Eqs.~(\ref{a8}) with Eqs.~(\ref{a3}). 
Although the equations for the diagonal matrix 
elements are the same, the equations for 
the off-diagonal matrix element are not. The difference is in the 
dissipative term, which includes now the detector tunneling rate 
$\gamma_L$. It is easy to trace its origin.  
In accordance with the Bloch 
equations the dissipation of the nondiagonal density-matrix elements 
$\bar\sigma_{bc}$ is the half of all possible decay rates of each 
of the states ($|b\rangle$ and $|c\rangle$). In the presence 
of the detector, the state $|c\rangle$, Fig. 2, has  
an additional decay channel, corresponding to the possibility for an 
electron to enter the detector. Although  
the time which an electron spends in the detector tends to zero, 
so that the related detector state does not distort the measured 
system, the {\em possibility}  
of such a process may influence the measured current 
very drastically. Indeed, if $\gamma_L\gg \Gamma_{L,R},\Omega$, 
one finds from Eq.~(\ref{a8d}) that $\bar\sigma_{bc}\to 0$ 
and therefore $I_S=e\Gamma_R\bar\sigma_{bc}\to 0$. 

The described strong damping of the nondiagonal density-matrix 
elements takes place only when the detector can {\em distinguish} 
a particular dot occupied by an electron. 
Such an ``observation'' effect disappears if   
$\tilde E_F^L<E_0+U_2$. In this case an electron cannot 
enter the detector no matter which of the dots of the measured 
system is occupied. Then the additional decay channel for
the state $|c\rangle$ is blocked and 
Eq.~(\ref{a8d}) coincides with 
Eq.~(\ref{a3d}), i.e. the measured current remains undistorted, 
$I_S=I_S^{(0)}$, Eq.~(\ref{a7}). Such a peculiar dependence of the 
measure current $I_S$ on $\tilde E_F^L$ is shown in Fig. 3.
\begin{figure}
\vspace{0.2cm}
\hspace{0.4cm}
\epsfxsize=6cm
\epsfysize=4cm
\epsffile{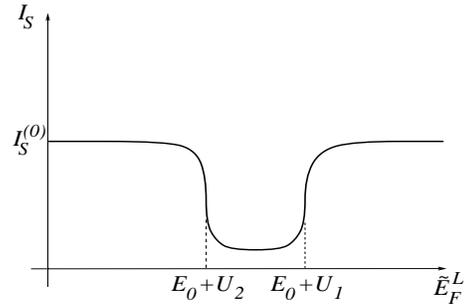}
\caption{Maximal current in the double dot structure ($E_1=E_2$)
as a function of the Fermi energy of the 
left reservoir adjacent to the detector.} 
\label{fig3}
\end{figure} 

Our analysis demonstrates that the 
disappearance of the non-diagonal matrix elements during 
the measurement process is 
attributed to the dissipation term. The latter is always generated 
by the Schr\"odinger equation by tracing out the continuum spectrum
states\cite{gp}.  
It is therefore no need to introduce an additional projection postulate
to describe the measurement process.\cite{zur,fre}. Most interesting aspect,
which reveals our analysis is that the 
reduction postulate is not necessary even for  
interaction-free (negative-result) measurement. 

Although the negative-result measurement which influences the measured
system without actual interaction is similar to  
the famous EPR paradox, some features 
are different. First, the above 
``observation'' effect appears as
a stationary state phenomenon. 
Second, we do not need any special initial correlations 
between electrons in the detector and the measured system. 
It leads to a possibility of influencing the measured 
current by switching the detector on (or off). Such a process which 
has a relaxation time $\sim 1/\gamma$, 
can also be studied using the same rate equations (\ref{a6}).
Most interesting problem 
would arise when the distance between the detector and the 
measured system is larger then $c/\gamma$, so that   
the density-matrix reduction, generated by 
the negative result measurement 
might contradict the relativistic requirement.  

Finally we like to mention that a similar description 
of the negative result measurement 
can be used for the optical experiment of Kwiat {\em et al.}\cite{kwiat}. 
In this experiment the 
non-observance of the ``object'' modifies the interference of the photon. 
In terms of our description it corresponds to the collapse of 
the photon density-matrix due to 
the {\em possibility} of the photon absorption by the ``object''.
Notice that the ``object'' would play a role of our  
``detector'' for a constant flux of photons.  
The time which takes such a collapse 
due to the non-observance of the object  
is therefore $\sim 1/{\bar\gamma}$, where $\bar\gamma$ 
is the photon absorption width. One thus can 
investigate a possible contradiction between 
the density-matrix reduction rate and the relativistic 
requirement also in 
the optical negative-result measurements.   

Parts of this work were done while the author stayed at 
University of Trento, Trento, Italy and TRIUMF, Vancouver, Canada. I thank 
these institutions for their hospitality. I am also  
grateful to A. Korotkov for very important comments, 
related to this work.


\begin{references}
\bibitem{neu} J. von Neumann, {\em Mathematische Grundlagen der Quantentheorie}
(Springer, Berlin, 1931). The statistical mixture means that the measured 
system is actually in one of its states, although there is no way to 
determine in which particular state the system is.
\bibitem{zur} W.H. Zurek, Physics Today {\bf 44}, No. 10, 36 (1991); 
{\em ibid}, {\bf 46}, No. 4, 13 (1993). 
\bibitem{renninger} M. Renninger, Z. Phys. {\bf 158}, 417 (1960); 
R.H. Dicke, Am. J. Phys. {\bf 49}, 925 (1981);  
A. Elitzur and L. Vaidman, Found. Phys. {\bf 23}, 987 (1993). 
\bibitem{vdr} N.C. van der Vaart {\em et al.},  
Phys.\ Rev.\ Lett. {\bf 74}, 4702 (1995); F.R. Waugh {\em et al.},
Phys.\ Rev.\ Lett. {\bf 75}, 4702 (1995).
\bibitem{pep} M. Field {\em et al.},
Phys. Rev. Lett. {\bf 70}, 1311 (1993); 
L.W. Molenkamp {\em et al.},   
Phys. Rev. Lett. {\bf 75}, 4282 (1995).
\bibitem{kwiat} P. Kwiat {\em et al.},
Phys.\ Rev.\ Lett. {\bf 74}, 4763 (1995).
\bibitem{bg} I. Bar-Joseph and S.A. Gurvitz, Phys. Rev B{\bf 44}, 3332 (1991).
\bibitem{glp} S.A. Gurvitz, H.J. Lipkin, and Ya.S. Prager, 
Mod.\ Phys.\ Lett.\ B {\bf 8}, 1377 (1994); Phys. Lett. A {\bf 212}, 91 (1996).
\bibitem{gp} S.A. Gurvitz and Ya.S. Prager, Phys. Rev. B{\bf 53}, 15932 (1996).
\bibitem{naz} Yu.V. Nazarov, Physica B {\bf 189}, 57 (1993),
T.H. Stoof and Yu.V. Nazarov, Phys. Rev. B {\bf 55}, 1050 (1996).
\bibitem{fre} V. Frerichs and A. Schenzle, 
Phys. Rev. A{\bf 44}, 1962 (1991); A. Beige and G.C Hegerfeldt, 
Phys. Rev. A{\bf 53}, 51 (1996).
\end{references}
\end{document}